\documentclass{PoS}
\usepackage{epsfig}
\usepackage{graphicx}
\usepackage{longtable}

\newcommand{\solm}{M$_{\odot}$\ }

\title{Nuclear Activity and the Conditions of Star-formation at the Galactic Center}
\ShortTitle{The Center of the Milky Way}

\author{\speaker{Andreas Eckart}\\
I. Physikalisches Institut der Universit\"at zu K\"oln, Z\"ulpicher Str. 77, D-50937 K\"oln, Germany;
and Max-Planck-Institut f\"ur Radioastronomie, Auf dem H\"ugel 69, D-53121 Bonn, Germany;
E-mail: \email{eckart@ph1.uni-koeln.de}}

\author{M. Valencia-S.\\
I. Physikalisches Institut der Universit\"at zu K\"oln, Z\"ulpicher Str. 77, D-50937 K\"oln, Germany;
}
\author{B. Shahzamanian, M. Zajacek, L. Moser, M. Parsa\\
I. Physikalisches Institut der Universit\"at zu K\"oln, Z\"ulpicher Str. 77, D-50937 K\"oln, Germany;
and Max-Planck-Institut f\"ur Radioastronomie, Auf dem H\"ugel 69, D-53121 Bonn, Germany;
}
\author{M. Subroweit, F. Peissker, N. Sabha, M. Horrobin, C. Straubmeier\\
I. Physikalisches Institut der Universit\"at zu K\"oln, Z\"ulpicher Str. 77, D-50937 K\"oln, Germany;
}
\author{A. Borkar, D. Kunneriath, V. Karas\\
Astronomical Institute of the Academy of Sciences Prague, Bocni II 1401/1a, CZ-141 31 Praha 4, Czech Republic
}
\author{C. Rauch, S. Britzen, A. Zensus\\
Max-Planck-Institut f\"ur Radioastronomie, Auf dem H\"ugel 69, D-53121 Bonn, Germany;
}
\author{M. Garc\'{i}a-Mar\'{i}n\\
European Space Agency (ESA/STScI), 3700 San Martin Drive, 
Baltimore, MD 21218, USA
}

\abstract{
The Galactic Center is the closest galactic nucleus that can be studied with unprecedented
angular resolution and sensitivity. We summarize recent basic observational results on Sagittarius~A* 
and the conditions for star formation in the central stellar cluster.
We cover results from the radio, infrared, and X-ray domain and include results from simulation as well.
From (sub-)mm and near-infrared variability and near-infrared polarization data we find that 
the SgrA* system (supermassive black hole spin, a potential temporary 
accretion disk and/or outflow) is well ordered in its geometrical orientation and in its 
emission process that we assume to reflect the 
accretion process onto the supermassive black hole (SMBH).
}
\FullConference{Frontier Research in Astrophysics - II\\
		23-28 May 2016\\
		Mondello (Palermo), Italy}

\begin{document}


\section{Introduction}

Sagittarius~A* (Sgr~A*) at the center of our Galaxy is a highly variable 
near-infrared (NIR) and X-ray source which is associated with a 
$4 \times 10^{6}$ \solm supermassive central black hole (see discussion in Eckart et al. 2017).
This region allows us in an unprecedented way to study at the 
same time, the nuclear activity associated with accretion onto 
the supermassive black hole, and the possibility for 
star formation in its vicinity.
Accretion and star formation models must take into account 
the presence (or influence) of
several dusty sources in the central 1~pc region
(Meyer et al. 2014, Eckart et al. 2013).
The fast motion of one of those infrared excess sources was discovered
by Gillessen et al. (2012).
They interpret it as a core-less (i.e. no star at its center) gas and dust cloud
approaching SgrA* on an elliptical orbit.
Eckart et al. (2013ab) present K$_s$-band identifications (from VLT and Keck data)
and proper motions of this DSO
\footnote{DSO stands for Dusty S-cluster Object. 
The source is also called G2 in the literature.
The S-cluster is the cluster of high velocity 
stars surrounding SgrA*; see Eckart\&Genzel (1997). 
The DSO passed by the central supermassive black hole of the Milky Way
SgrA* within $\sim$150-160~AU.}.
Here we present a brief description of the accretion process acting onto SgrA*
from a statistical analysis of emission at different wavelengths. We also explore the
geometrical properties of SgrA* accretion flow and of the DSO via infrared polarimetry
(Shahzamanian et al. 2015, 2016).

\section{SgrA* Emission Process}

\subsection{Radio/sub-mm monitoring of SgrA*}

\begin{figure}[!ht]
\begin{center}
\includegraphics[width=13cm]{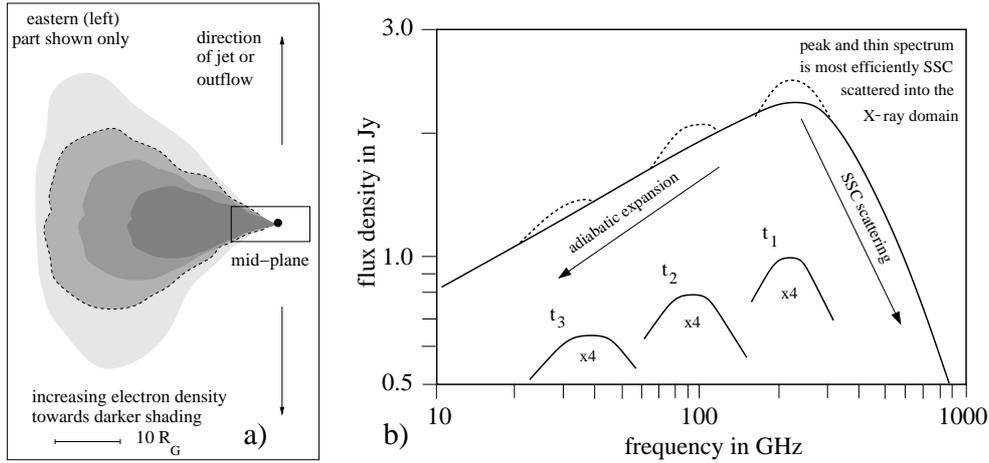}
\caption{
a) Sketch of the density distribution over the azimuthal section across a 
Sgr~A* black hole accretion torus. 
This depicts the result from relativistic magneto hydrodynamic modeling
following Moscibrodzka (2009).
The mid-plane is located within the box, to the top and bottom one finds
the jet/wind region.
b) Depicting schematically the spectral properties of the adiabatic expansion process.
} \label{fig1}
\end{center}
\end{figure}

Sgr A* undergoes radio millimeter and sub-millimeter variability.
Most of this variable flux density is thought to arise from the very central region of the accretion
flow onto the supermassive black hole (Fig.~\ref{fig1}a; e.g. Moscibrodzka et al. 2009, 2013).
In Subroweit et al. (2016), we report on a detailed statistical analysis of the sub-millimeter 345~GHz 
and radio 100~GHz flux density distribution of Sgr~A*. 
The millimeter radio observations
were carried out using the Australia Telescope Compact Array (ATCA) in the years between 2010 and 2014 (Borkar et al. 2016).
In the sub-mm wavelength domain, we used data obtained from several Large Apex Bolometer Camera (LABOCA) observing 
campaigns in the years 2008 to 2014. In addition to this, we used 
literature data covering the years from 2004 to 2009, at frequencies very close to 345~GHz.
The ATCA observations resulted in the detection of six bright flux density excursions of Sgr A*.
During these events the fluxes varied between 0.5 and 1.0 Jy. These events lasted for typically 1.5 to 3 hours.

\begin{figure}[!ht]
\begin{center}
\includegraphics[width=13cm]{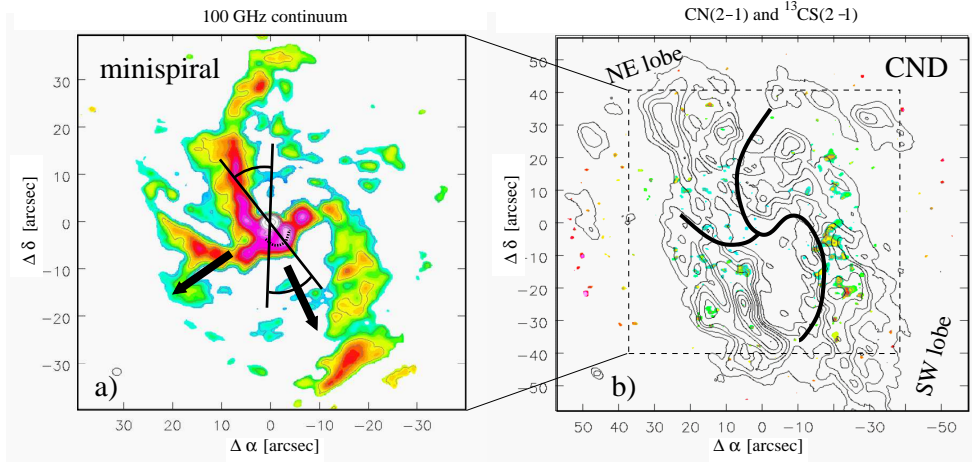}
\caption{
a) ALMA continuum image of the minispiral surounding SgrA*
(Moser et al. 2016)
and the NIR polarization properties of SgrA*.
The conus shows the range over which the polarization angles 
typically vary. The thick black arrows point into 
jet directions discussed in the literature (see Shahzamanian et al. 2015).
b) 
The CND in molecular line emission. In contours, 
as shown by Martin et al. (2012), obtained with the SMA
Sub-Millimeter Array at a resolution of (4.0''$\times$2.6'').
In color we show the $^{13}$CS(2-1) line emission as shown by 
Moser et al. (2016) with flux densities around 0.2 Jy beam$^{-1}$s$^{-1}$
at a resolution of (1.9''$\times$1.6'').
} \label{fig2}
\end{center}
\end{figure}

Moser et al. (2016) report serendipitous detections of line emission 
with the Atacama Large Millimeter/submillimeter Array (ALMA).
The observations were carried out within the central parsec around 
SgrA* at a resolution of up to 0.5''.
From the 100~GHz continuum and line flux densities
(example maps are shown in  Fig.~\ref{fig2}ab),
Moser et al. (2016) obtained temperature and density 
estimates towards different source components.
The spectral index (S$\propto$$\nu$$^{\alpha}$) 
of the synchrotron emitting material in the immediate surroundings 
of Sgr A* is around $\alpha$=0.5 at 100 - 250~GHz and 
around $\alpha$=0.0 at 230 - 340~GHz interval.
From earlier interferometric and single dish sub-mm observations 
and modeling we know that the overall spectrum drops strongly towards 
the far-infrared at frequencies above about of 350 GHz 
(Marrone 2006,  Marrone et al. 2006a,b, Eckart 2012).
Other Galactic Center regions emitting continuum radiation indicate contributions 
from Bremsstrahlung (around $\alpha$=-0.1) 
and contributions of cold dust (with flatter/inverted spectral indices).

Subroweit et al. (2016) and Borkar et al. (2016)
show that the variable millimeter flux density of SgrA* can be explained 
via the adiabatically expanding plasmon model.
For earlier events Yusef-Zadeh et al. (2006) explained 
the flaring activity of Sagittarius~A* 
at 43 and 22~GHz as evidence for expanding blob of hot plasma.
This phenomenological model was soon developed with much more solid 
dynamical base and an MHD model for the general picture of plasmoid 
ejection and associated flare was proposed by Yuan et al. (2009).
Most recently, the model was further developed by calculating the 
detailed radiation of the ejected blob based on dynamics and compared 
with flare observations of Sgr~A* (Li et al. 2017).
In Fig.~\ref{fig1}b we show schematically how a sub-mm flare 
component evolves for three different consecutive times. 
As a result of the adiabatic expansion, the peak flux decreases 
and the spectral width of the component increases.
These physical parameters derived from the modelling 
imply that the expanding source components 
responsible for the flux density variability of SgrA*,
are either confined to the immediate
vicinity of Sgr A* or they have a bulk motion greater 
than their expansion velocity.
They may be generated in the temporary disk or corona component of SgrA*.

The flux-density variations were not obviously affected by the 
approach or the flyby of the DSO. 
This implies that the source must be very compact and does not
have a large bow shock.

Subroweit et al. (2016) show that the APEX and ATCA light curves can be described by a shifted power law.
The dependency can be written as  p(x)$\propto$(x - s)$^{-\alpha}$ with $\alpha$$\sim$4 
(APEX: $\alpha$ = 4.0 $\pm$ 1.7, ATCA: $\alpha$ = 4.7 $\pm$ 0.8). 
A similar power law index was found investigating near-infrared (NIR) light curves (Witzel et al. 2012).
The value measured in the optically thin NIR parts (Witzel et al. 2012) has been obtained 
with very good statistics from a large number of lightcurves. The 100~GHz and 345~GHz 
(sub-)mm values shown in Fig.~\ref{fig3}a reflect this NIR value very well.
The data points with vertical ranges in Fig.~\ref{fig3}a are taken from Subroweit et al. (2016) and 
Witzel et al. (2012). The horizontal bar reflects the range of synchrotron turn-over 
frequencies for the case of a synchrotron source with a magnetic field of less than 30~G, peak flux densities of 
less than 2~Jy, source sizes of less than 2 Schwarschild radii, and an optically thin spectral
index of -0.7 under the assumption that the NIR is still optically thin synchrotron 
radiation and the X-ray domain stems from Synchrotron Self-Compton (SSC) radiation (Fig.6, Eckart et al. 2012).
This region marks where the turnover frequencies of probably most of the bright radio flares are located.
These flares then expand adiabatically into the radio, provid the optically thin synchrotron emission in the NIR,
and SSC-scatter into the X-ray domain. 

The overall findings support our preferred flare emission models in which
the adiabatically expanding plasmons dominate the radio - sub-mm short term 
variability (see Fig.~\ref{fig1}ab).
When they are most compact they are responsible for the variable optically 
thin NIR emission and produce variable X-ray emission via the SSC effect.

In the model the flux density variations in all wavelength bands stem from the same 
source components and statistically
they represent a single state red noise process. 
We find that a typical expansion velocity is of the order of 0.01~c and the dominant 
portion of variable source components are 
generated with a peak turn-over frequency in the range of about 300-400~GHz.
The contribution of higher frequency emission flares that expand faster or 
flares that are born with turnover frequencies around 100~GHz is minimal.
These stonger flares are stronger and superimposed on the source emission fluctuations that consist
of the contributions of more extended and (on short times scales) less variable 
 source components.

\subsection{Variable radio structure of SgrA*}

In order to determine the nature of some of these (sub-)millimeter flux density excursions
we performed near-infrared triggered mm-VLBI observations of the radio source Sgr A* at 43~GHz
(Rauch et al. 2016).
The compact radio and near-infrared source SgrA* associated with the supermassive black hole in
the Galactic center was observed in the NIR and mm wavelength range.
During a global multiwavelength campaign in May 2012
and based on a NIR flare, we triggered a global Very Long Baseline Array (VLBA) campaign.
Here the observations were carried out with the VLBA at 7 mm (43~GHz) for 6 h each day.
As a result we found that typically the total 43~GHz flux density of Sgr A* shows only minor variations ($\le$0.06~Jy)
during its quiescent phases on a daily basis.

\begin{figure}[!ht]
\begin{center}
\includegraphics[width=13cm]{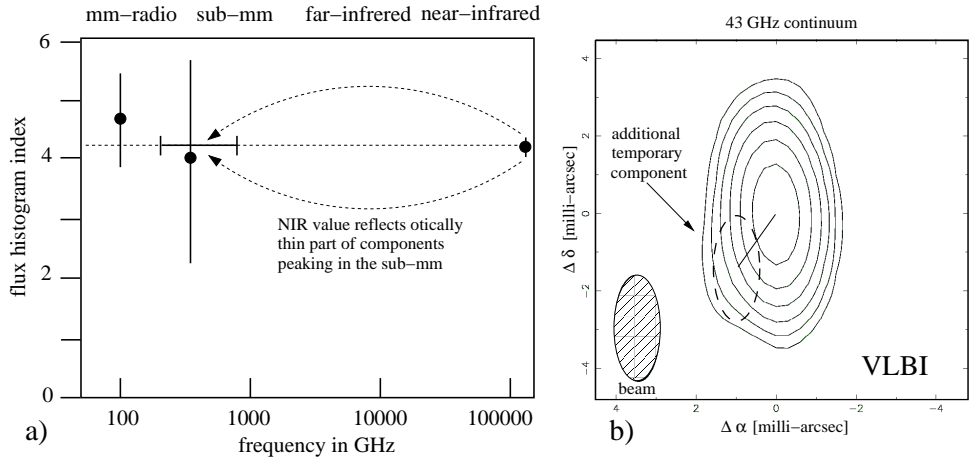}
\caption{
a) Slope of the flare amplitude histograms in different wavelength regimes.
b) Right hand circular polarization map of the 43~GHz mm-emission of Sgr A* on 
May 17, 2012 (8:00-10:00 h UT). The clean beam is shown as a 
line shaded insert.
The contour levels are 1.73\%, 3.46\%, 6.93\%, 13.9\%, 27.7\%, 
and 55.4\% of the peak flux density of 1.5 Jy/beam (see Rauch et al. 2016 for details).
} \label{fig3}
\end{center}
\end{figure}

We triggered, however, the VLBI observations by a NIR flare observed at the Very Large Telescope (VLT).
During this campaign we measured a NIR flare on May 17, 2012, that preceeded a 43~GHz flare 
of 0.22 Jy by about 4.5 hours.
This time delay is consistent with expected values due to adiabatic expansion. 
Since we were carrying out imaging VLBI observations we could follow the SgrA* 
source structure during the flare.  Close to the  peak time of the radio flare, Sgr A*
showed a secondary radio component located at about 1.5 mas toward the southeast of the core position. 
In Fig.~\ref{fig3}b we show a map at the time of the mm-flare event with the secondary component indicated.

A source component at that distance from the core would be consistent with relativistic 
bulk motions of the expanding component.  Hence, it could be a jet or an outflow component.
The event can be explained by a bluk motion at the speed 
of almost half the speed of light.
While variations in the intrinsic source structure have been observed several times
(see e.g. Bower et al. 2014),
we cannot exclude that radio sub-millimeter scintillations
have been involved in the observed effects.

\subsection{X-ray activity of SgrA*}

With a black hole mass of 4 million solar masses
the bolometric luminosity of SgrA* of $L_{bol}\sim~10^{36}~erg/s$ 
is lower than its Eddington luminosity of 
$L_{Edd} = 3 \times 10^{44} erg/s$
(Yuan et al. 2003).
Such a low luminosity points at radiatively inefficient accretion 
flow models (e.g. advection dominated accretion flows - ADAF; 
Narayan et al. 1998) or models that involve a jet-disk system.
SgrA* is not only variable in the radio/sub-mm and infrared domain
but also in the X-ray domain. In addition to a 1'' diameter 
quiescent Bremsstrahlung component, there are X-ray flaring events 
from Sgr A* - first discovered by Baganoff et al. (2001).
Typically one finds one bright flare per day (Neilsen et al. 2013) 
about 10 times the quiescent (2-8 keV) luminosity of SgrA* of about
$3.6 \times 10^{33} erg/s$
(Baganoff et al. 2003; Nowak et al. 2012).
Flares of up to 160 times the quiescent level have been 
reported (Porquet et al. 2003, 2008; Nowak et al. 2012). 

In the NIR we observe optically thin synchrotron radiation 
emission with a spectral index of about $\alpha$=-0.7
($S_{\nu} \propto \nu^{\alpha}$)
and a flare amplitude statistics as described by (Witzel et al. 2012).
The least demanding requirements to produce the X-ray flux densities
appears to hint at a radiation process that involves Comptonization. 
Here, moderate volume densities ($10^6 cm^{-3}$)
of low relativistic electrons with $\gamma$$\sim 10^3$ are required
(e.g. Eckart et al. 2004 2008, Yusef-Zadeh et al. 2006).

Mossoux et al. (2016) observed a total of seven NIR flares, three of which 
have XMM-Newton detected X-ray counterparts. 
Since the flaring rate during this campaign was fully consistent with
that of the 2012 Chandra campaign one must conclude that 
no significantly increased activity was detected 
during/close to the pericenter passage of the DSO.

\begin{table}[htb]
\caption{Conditions for star-formation in the central parsec}
\begin{center}
\begin{tabular}{ccccc}\hline \hline
source component                              & density & temperature & estimated mass  \\  
                                              & cm$^{-3}$& K& \solm \\ \hline
                                              & based on observations & & \\ \hline
Circum Nuclear Disk (CND)$^{3,4,5}$           & 10$^{5}$-10$^{6}$  & 150-450 &  $10^{4-5}$ \\ 
mini-spiral (plasma)$^{2,1}$                  & $\le$10$^{4.0}$             & 6000 & $\le$3 \\
mini-spiral (molecular gas/dust)$^{2,6,7,8}$ & 10$^{4.5}$-10$^{6.5}$  & 100-300 & 5-50 \\
center within CND$^{1,9}$                     & $\le$10$^{2}$  & 60 & $<$0.05 \\
without mini-spiral                            &            &     &    \\
central arcsecond$^{1,9}$                     &$<$10$^{4}$  & $>$120 & $<$3$\times$10$^{-4}$ \\ \hline
                                              & model requirements & & \\ \hline
gas and dust disk                             & 10$^{8}$-10$^{13}$  & 500-1000 &(2-4)$\times 10^4$ \\
in Nayakshin's model$^{10,11}$                &            &     &    \\
gas and dust streamer                         & 10$^{5}$-10$^{11}$  & 10-50& 100 \\
in Jalali's model$^{12}$                      & & \\
\hline \hline
\end{tabular}
\label{Tab1}
\end{center}
Literature key: 1) Shcherbakov et al. 2010;
2) Moser et al. 2016;
3) Wright et al. 2001;
4) Christopher et al.  2005;
(with clumps exceeding densities of 
10$^{7-8}$cm$^{-3}$ and temperatures of 50~K or less);
5) Mills et al. 2013;
6) Kunneriath et al. 2012;
7) Jackson et al. 1993;
8) Latvakoski et al. 1999;
9) Rozanska et al. 2014;
10) Nayakshin, Cuadra \& Springel 2007 (with critical density for 
star formation above 10$^{13}$cm$^{-3}$);
11) Nayakshin \& Cuadra 2005;
12) Jalali et al. 2014
 (with critical density for star formation above 10$^{11}$cm$^{-3}$).
\end{table}

\subsection{Infrared Polarimetry of SgrA*}

Polarized radiation gives information on the emission process and the geometry of the
emitting source components. 
In the NIR at a wavelength of 2.2$\mu$m, SgrA* is strongly linearly polarized.
Observations have been carried out using the adaptive optics
instrument NACO at the VLT UT4 in the infrared K$_s$-band (2.00$\mu$m - 2.36$\mu$m) from 2004 to 2012.
Several polarized flares were observed during these years, allowing us to study the statistical
properties of linearly polarized NIR light from Sgr A*.
In Fig.~\ref{fig2}a we show how the source is embedded within the mini-spiral
(shown in radio continuum emission) and depict the NIR polarization properties.
At flux density excursions above 5mJy (Shahzamanian et al. 2015)
the number density histogram for polarized flare fluxes 
has an exponent $dN/dS$$\sim$4, close to the slope of the
single state power-law distribution that describes the total power flare flux distribution
(Witzel et al. 2012).
The polarization degrees are typically around 20\%
at a preferred polarization angle of 13$^\circ$$\pm$15$^\circ$.
These observational facts imply that the geometry and energetics of the accretion process 
within the SgrA* system (black hole, wind or jet, and temporary disk or mid-plane as 
found in simulations; see Fig - see Fig.~\ref{fig1}a)
are rather stable.

\section{The Dusty S-cluster Object}

In Peissker et al. (2016) 
we summarize our monitoring of SgrA* and the Dusty S-Cluster Object (DSO) in the
near-infrared using VLT SINFONI in the years between 2006 and 2015.
The faint DSO was found in 2011 on its way towards the supermassive black hole 
at the center of our Milky Way. The object was primarily tracked in the thermal NIR 
L'-band and in its Br$\gamma$-line emission in the NIR K-band.
The near-infrared Ks-band continuum emission of the DSO was first reported by
Eckart et al. (2013) using NACO at the ESO VLT. It was then also confirmed using 
the NIRC system at the Keck telescope (Eckart et al. 2014) and SINFONI at the VLT (Eckart et al. 2015).
The clear transit between red- and blue-shifted line emission (Valencia et al. 2015), and the 
fact the the DSO remained a very compact continuum and recombination line emission source
(Valencia et al. 2015,  Witzel et al. 2014) clearly shows that it did not disintegrate as previously expected  
(e.g.  Gillessen et al. 2012, Pfuhl et al. 2015, Schartmann et al. 2015).
Using SINFONI data with high image quality Peissker et al. (2016) now show that 
the DSO is found as a compact Ks-band line and continuum emitter before and after periapse.

\begin{figure}[!ht]
\begin{center}
\includegraphics[width=13cm]{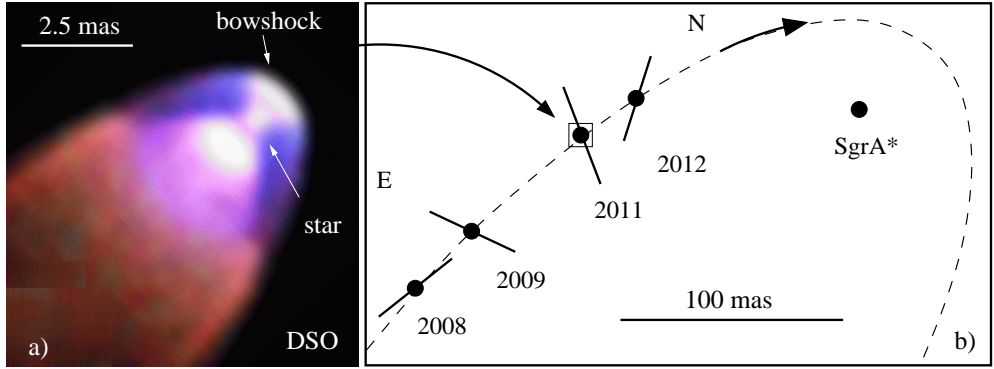}
\caption{
a) An RGB image of the source model of the DSO. 
While the DSO passes through the medium close to SgrA*
it develops a compact bow-shock structure
(Shahzamanian et al. 2016 and Zajacek, Karas \& Eckart 2014,
Zajacek et al. 2016).
b) Polarization properties of the DSO passing by
the SgrA* black hole. The polarization degrees are of the order of
30\%, the polarization angle varies as shown in the figure (Shahzamanian et al. 2016).
} \label{fig4}
\end{center}
\end{figure}

In Shahzamanian et al. (2016)
we investigate the near-infrared continuum of the DSO in more detail.
For the first time, Shahzamanian et al. (2016)
use the near-infrared polarimetric imaging data to
determine the polarization of this source.
From the data between 2008 and 2012 Shahzamanian et al. (2016) conclude from a 
robust significance analysis that the DSO is an intrinsically polarized source (> 20\%).
It is indicated that the polarization angle varied while it travelled towards
the periapse on its orbit around  SgrA*.
Also, based on the infrared excess of $K_s - L > 3$, the authors conclude that
the DSO might be a dust-enshrouded young star with a compact bow shock forming while it approaches 
the supermassive black hole (see Fig.~\ref{fig4}a).
The high polarization degree suggests that the scattering material is arranged in a 
non-spherical geometry (see e.g. Zajacek, Karas \& Eckart 2014, Zajacek et al. 2016).
Shahzamanian et al. (2016) model the DSO as a combination of a bow shock
and a  bipolar wind of the central star. 
This model is consistent with the DSO's total flux density and the polarization degree. 
The model naturally explains also the varying polarization angle 
as being due to intrinsic changes of the source structure, i.e. orientation of the dusty source 
with respect to the orbit (see Fig.~\ref{fig4}b).
In addition, an interaction of the DSO with the ambient medium may also influence the polarization angle.

\section{Conditions of Star Formation}

Luminous He-stars in the central few arcseconds (Krabbe et al. 1991 Ghez et al. 2003), 
the indications for young stars based on colors (Buchholz, Sch\"odel, Eckart 2009),
as well as the presence of several dusty objects like the DSO 
(Meyer et al. 2014, Eckart et al. 2013)
suggest that star formation is permanently, or at least episodically,
taking place in the central parsec - in the immediate vicinity of the supermassive black hole SgrA*.
In Table~\ref{Tab1} we list the gas densities and temperatures, as well as the integrated masses for different regions in the 
central parsec. This table shows that only the CND (Fig.~\ref{fig2}b) harbor dense gas entities 
that may give rise to star formation if a suitable trigger is applied, 
i.e. additional compression that brings the density above the Jeans density to initiate the collapse to a star.
However, for the central few arseconds other mechanisms have to be invoked.
It is often assumed that the strong gravitational forces close to SgrA* disrupt
dense  clouds and prevent classical star formation.
However, young stars are observed all over the entire nuclear star cluster.
While the massive young He-stars, 
most of which are located in a disk (Levin \& Beloborodov 2003, Bartko et al. 2008),
have most likely been generated in a massive gaseous disk 
(Nayakshin, Cuadra \& Springel 2007, Nayakshin \& Cuadra 2005),
the lower mass stars may have been formed in a different way.
Jalali et al. (2014) have shown that
molecular clumps of typically 100 \solm at a radius of less than 0.2 parsec 
may be subject to black-hole supported star formation due to the orbital compression the
clumps experience while they go through periapse,  if their temperature is similar to that found in 
CND clumps. Through dissipative cloud-cloud collisions in the CND the clumps lose angular momentum 
and can get to pericenter passages close to the supermassive black hole.
In Table~\ref{Tab1} we also list the densities, temperatures and masses  covered by the corresponding 
model calculations. For the Nayakshin case the temperature is higher for a more massive 
star forming entity,  resulting also in more massive stars.
The Jalali-mechanism can result in low and intermediate mass star formation that may 
very well explain the presence of DSO like objects, i.e. compact, dust enshrouded young stars.

\section{Summary}
We have presented results of recent monitoring programs covering the radio to X-ray regime 
for the Galactic Center as the closest galactic nucleus available.
The radio/sub-mm monitoring indicates that the bright 0.1 Jy to a few Jy flare emission of SgrA* 
can be explained by a simple adiabatic expansion model, in which flares are born
within turnover frequencies in the 300-400~GHz range. To first order they are 
of similar nature, and are consistent with the spectral index and amplitude index information 
that can be derived from near-infrared and (sub)mm observations.
The results from triggered VLBA observations indicate that some of the variable radio components 
may indeed be traveling outwards as part of a jet or wind.
The ordered geometry required for such a scenario is supported by the results from NIR polarization measurements.
In general, the flare emission from the radio, near-infrared to X-ray emission may be linked 
through a synchrotron-self-Compton mechanism.
Simulations and observations indicate that ongoing star formation at the Galactic Center can be sustained 
by black-hole supported star formation, either via the formation of a massive star forming disk or by 
triggering the compression of dense molecular clumps from the CND during periapse on in-falling orbits.
The Dusty S-cluster Object may be an example of a young stellar low-mass system.

\section*{Acknowledgements}
We received funding
from the European Union Seventh Framework Program
(FP7/2013-2017) under grant agreement no 312789 - Strong
gravity: Probing Strong Gravity by Black Holes Across the
Range of Masses. 
This work was supported in part by the
Deutsche Forschungsgemeinschaft (DFG) via the Cologne
Bonn Graduate School (BCGS), the Max Planck Society
through the International Max Planck Research School
(IMPRS) for Astronomy and Astrophysics, as well as special
funds through the University of Cologne and
SFB 956 – Conditions and Impact of Star Formation. M. Zajacek, M. Parsa and
B. Shahzamanian are members of the IMPRS. Part of this
work was supported by fruitful discussions with members of
the European Union funded COST Action MP0905: Black
Holes in a Violent Universe and the Czech Science Foundation
- DFG collaboration (No. 13-00070J).

\end{document}